# Matter-Antimatter Propulsion via
# QFT Effects from Parallel Electric and Magnetic Fields

**Gerald B. Cleaver**
Dept. of Physics & Center for Astrophysics, Space Physics & Engineering Research
Baylor University, Waco, Texas 76798-7316
Gerald_Cleaver@baylor.edu

## ABSTRACT

Matter/antimatter (MAM) pair production from the vacuum through intense electric fields has been investigated theoretically for nearly a century [1]. This presentation will review this history and will examine proposals of MAM for intra-solar system and interstellar propulsion systems. The quantum mechanical foundation of MAM production was developed by F. Sauter et al. in the 1930's and then placed on a sound quantum electromagnetics (QED) basis by J. Schwinger in 1951. Pair production occurs when the electric field strength $E_0$ is above the critical value at which the fields become non-linear with self-interactions (known as the Schwinger limit). As the energy density of lasers approach the critical strength of $E_0 \sim 10^{16}$ V/cm, the feasibility and functionality of electron-positron pair production has received growing interest. Current laser intensities are approaching within 1 order of magnitude of the Schwinger limit.

Physical processes for lowering the critical energy density below the Schwinger limit (and simultaneously enhancing the pair production above the Schwinger limit) through additional quantum mechanical effects have been explored. One under study at the U. of Connecticut and the U. of Duisburg-Essen is pulsation of inhomogeneous electric fields within a carrier wave. Another is via enhancement of quantum effects by addition of a magnetic field **B** parallel to the electric field **E**. Magnetic field enhancement to quark/anti-quark production through chiral symmetry breaking effects in quantum chromodynamics (QCD) was investigated theoretically by J. Preskill at Caltech in the 1980's. S. Pyo and D. Page showed in 2007 that parallel magnetic fields also enhance electron/positron production via an analogous QED effect, with enhancement going predominantly as a linear function of $B_0/E_0$, Particle/antiparticle pair production as a highly efficient fuel source for intra solar system and interstellar propulsion was proposed by D. Crow in 1983. The viability of this method of propulsion will be studied, especially from the parallel electric and magnetic field approach.

---

[1] Particle/anti-particle pair production does not (and cannot) take energy from the spacetime vacuum. Rather the energy is drawn from the external electric (and magnetic) fields. This process is very analogous to particle production near the event horizon of a black hole, which reduces the mass of the black hole accordingly. (The primary difference between the two processes is, while both particle and antiparticle are produced from a virtual pair by the electric (and magnetic) fields, only one particle in an initially virtual pair escapes from a black hole (as Hawking radiation) and the antiparticle is captured by the black hole.)

# INTRODUCTION

Matter-Antimatter (MAM) (a.k.a., particle/anti-particle) pair production from the spacetime vacuum through intense electric fields has been investigated for nearly a century. This paper reviews its history and examines proposals of MAM production for intra-solar system and interstellar propulsion systems. The quantum mechanical foundation of MAM pair production was developed by Fritz Sauter, Werner Heisenberg, and Hans Euler in the 1930's [1] and then placed on a sound quantum electrodynamics (QED) basis by Julian Schwinger in 1951 [2].

MAM production occurs when the electric field strength E is at or above the critical value $E_S$ (known as the Schwinger limit) at which the electromagnetic fields become non-linear with self-interactions. This non-linearity occurs when the energy within a Compton wavelength of a photon is equal to or greater than twice the rest mass of an electron. It corresponds to an electric field strength $E_S \equiv m^2c^3/(e\hbar) = 1.3 \times 10^{16}$ V/cm (equivalently, an electric field intensity $I_S = 2.1 \times 10^{29}$ W/cm$^2$).

In a vacuum, the classical Maxwell's equations are perfectly linear differential equations. This implies – by the superposition principle – that the sum of any two solutions to Maxwell's equations is yet another solution to Maxwell's equations. For example, if two beams of light interact linearly when aimed toward each, their electric fields simply add together and pass right through each other. In QED, however, non-linear photon–photon scattering becomes possible when the combined photon energy is large enough to spontaneously create virtual electron–positron pairs. When the average strength of an electric field is above $E_S$, the pair production rate (PPR) of charged particles per unit time and unit cross-section is found from the probability of quantum mechanical "tunnelling" of virtual MAM pairs from the Dirac sea into real particles.

(It should perhaps be emphasized that MAM production *does not (and cannot) steal energy from the spacetime vacuum*. Rather the energy is drawn from the external electric (and possibly magnetic) fields. The MAM production process is in many ways analogous to particle production near (on the outer side of) the event horizon of a black hole, which reduces the mass of the black hole accordingly. The primary difference between the two processes is, while both particle and antiparticle are produced from a virtual pair by the electromagnetic fields, only one particle in an initially virtual pair escapes from a black hole (as Hawking radiation) and the antiparticle is captured by the black hole.))

As the energy density of lasers approach the critical strength $E_S$, the feasibility and functionality of electron-positron pair production has received growing interest. Current laser intensities are approaching within 1 order of magnitude of this Schwinger limit. Examples are the X-ray free electron lasers at SLAC's the Linac Coherent Light Source and DESY's TESLA. Site four of the Extreme Light Infrastructure (ELI) Ultra-High Field Facility (UHFF) is planned for construction in eastern Europe around 2020 and should reach $E_S$. It will be composed of ten lasers concentrating 200 petawatts of power into a very narrow beam of $10^{-12}$ s pulses.

Physical processes for effectively lowering the critical energy density below the Schwinger limit (and simultaneously enhancing MAM PPR above the Schwinger limit) through additional quantum mechanical effects continue to be explored. Research teams at the U. of Connecticut and the U. of Duisburg-Essen are jointly examining critical energy field strength scale reduction/increased PPR via pulsation of inhomogeneous electric fields within a carrier wave [3].

Other investigations have focused on enhancement of quantum effects by addition of a magnetic field $\underline{B}$ parallel to the electric field $\underline{E}$. Magnetic field enhancement to quark/anti-quark production through chiral symmetry breaking effects in quantum chromodynamics (QCD) was investigated theoretically by John Preskill at Caltech in the 1980's [4]. S. Pyo Kim at the Kunsan National University and Don Page at the University of Alberta showed in 2007 that parallel magnetic fields also enhance electron/positron production via an analogous QED effect [5], with enhancement going predominantly as a linear function of B/E, the ratio of the magnitude of the magnetic and electric fields.

MAM production as a highly efficient fuel source for intra solar system and interstellar propulsion was proposed by Devon Crow in 1983 [6] and Robert Forward in 1985 [7]. The viability of this method of propulsion is considered below, especially with regard to the two PPR enhancement methods.

## BRIEF HISTORY OF MAM PRODUCTION

In 1928, British physicist P.A.M. Dirac showed that Einstein's relativity implied every type of particle has a corresponding antiparticle, with an identical mass, but opposite electric charge. Then in 1932, Carl Anderson at Caltech recorded discovery of a positively charged electron (i.e., positron) passing through a lead plate in a cloud chamber, for which he received the Nobel Prize in Physics. Two decades later in 1955, the antiproton was experimentally confirmed at Berkeley by Emilio Segre and Owen Chamberlain, earning them the 1959 Nobel Prize in Physics. Within one year, the antineutron was discovered at the Bevatron at Lawrence Berkeley Nation Lab by Bruce Cork and colleagues.

By 1995, researchers were using CERN's Low Energy Antiproton Ring (LEAR) to slow down antiprotons. They managed to pair positrons and antiprotons together, producing nine hydrogen anti-atoms, each lasting a mere 40 nanoseconds. Within three more years, CERN was producing approximately 2000 anti-hydrogen atoms per hour. Production rates of antimatter at CERN's Large Hadron Collider have steadily increased significantly since. (Likewise at Fermilab's Tevatron accelerator until its 2011 shut down).

## MAM AS PROPULSION SOURCE

MAM could be an ideal rocket fuel because all of the mass in MAM collisions can be converted into energy and used for thrust. MAM reactions produce 10 million times the energy produced by conventional chemical reactions used to fuel the space shuttle. It is 1,000 times more powerful than nuclear fission produced at a nuclear power plant & 300 times more powerful than the energy released by nuclear fusion. (Note however that should such an amount of antimatter be produced or collected, unless used for propulsion very soon thereafter, a secure means of long-term storage (i.e., magnetic confinement) would likely need to be devised. Antimatter must be kept separate from matter until a spacecraft needs more power, unless stored as anti-hydrogen. The alternative is for MAM to be created in situ and immediately emitted as propellant.

In 2000, NASA scientists announced early designs for a MAM engine that might be capable of fueling a spaceship for a trip to Mars using only a few milligram of MAM. In 2012, R. Keane and W.M. Zhang examined magnetic nozzle designs for charged pion emission from quark/antiquark collisions. Their study indicated that effective exhaust speeds ~ 0.7 c are feasible by optimizing nozzle geometry and magnetic field configuration using a magnetic field of order ~ 10 T. They also estimated an emission efficiency ~ 30% for MAM emission is obtainable for quark/antiquark pair production leading to pion emission and greater than 30% efficiency for electron/positron emission [8].

With MAM production a known technolgy [9], one might wonder what are the main hinderances to construction of MAM propulson systems. A primary issue is antimatter remains *the most expensive substance on Earth*. In 2000, it cost $62.5 trillion per microgram (equivalently, $1.75 quadrillion per ounce) of electron/positron pairs, with Fermilab able to produce only about 15 nanograms a year. However, the price of antimatter has continued to drop with each advancement in particle accelerator intensity and efficiency. CERN's LHC now produces about 1 microgram of antimatter (equivalently, $10^{21}$ electron/positron pairs) per 12 days at a cost of $200,000 or 1 milligram in about 12,000 days (that is, 30 years) at a cost of around two-hundred-billion dollars.

In its *Status of Antimatter* report, the NASA Glenn Research Center (www.nasa.gov/centers/glenn/technology/warp/antistat.html, dated 14 July 2015) concluded that for MAM to be a commercially viable fuel for travel within our solar system, "the price of antimatter would need to drop by about a factor of ten-thousand." Based on the rate of decline of antimatter production cost over the last 25 years, and its extrapolation into future decades, the NASA cost reduction goal may be obtainable within one to two decades from now. 2025 to 2035 was the time scale for MAM cost viability predicted by Crow and Forward in the 1980's.

Significantly more than just a few milligrams of MAM are required for interstellar travel, even to the closest star systems. Further, if planet reconnaissance or a landing mission is involved, additional MAM is needed to decelerate a spacecraft into the target star system. A spacecraft with a 100-ton payload designed for to cruising at 0.40 c is estimated to require the equivalent of 80 ocean supertankers full of MAM fuel [10]. (However, for somewhat lower cruise speed ~ 0.25 c, MAM requirements are dramatically lowered, but still remain extremely large [11].)

One possible solution to the extremely high cost of MAM production on earth is collecting MAM in space. In 2011, antiprotons were discovered by the international PAMELA (Payload for Antimatter/Matter Exploration and Light-nuclei Astrophysics) satellite to be trapped by Earth's magnetic field. The Alpha Magnetic Spectrometer on ISS is also able to detect, identify, and measure antiparticles in Earth orbit. Theoretical studies relatedly suggest that the magneto-spheres of much larger planets, like Jupiter, should have significantly more antiprotons than earth. Keane and Zhang point out that "if feasible, harvesting antimatter in space would completely bypass the obstacle of low energy efficiency when an accelerator is used to produce antimatter" [8].

# IN SITU MAM GENERATION

In addition, an ideal MAM propelled spacecraft should contain systems for both *collecting and generating* MAM, with creation especially as an emergency option if the stored antimatter leaks out of magnetic containment chambers or is annihilated prematurely by matter leaking in. Significant developments in both theoretical and engineering aspects of MAM production via strong localized electromagnetic fields have occurred in the last decade. For example, in [5] Sang Kim and Don Page derived the MAM production rate from a static plane-symmetric z-dependent electric field E(z): Consider a static plane-symmetric z-dependent electric field E(z) in the z-direction, with maximum value $E_0$ and of effective length L such that $E_0 L = ½ \int E(z)\, dz$. This arrangement allows pair production of a particle of mass m and charge q if $\varepsilon \equiv m/(qE_0 L) < 1$ or equivalently $E_0 > m/(qL)$ (in natural units of $G_N = c = \hbar = 1$). (Alternately, if we want a time varying field E(t) rather than a spatially varying field, replace $\varepsilon$ with $\varepsilon_T$, L with T, and dz with dt. (In that case, pair production occurs even with $\varepsilon_T > 1$, but is suppressed.)

In both the spatially-varying and time-varying processes, when $E_0$ is above the minimum value, MAM PPR of charged particles per unit time and unit cross-section can be computed from tunnelling of virtual pairs from the Dirac sea, where instantons determine the QM tunnelling probabilities. To leading WKB order, for a "Sauter" electric field of the form $E(z) = E_0 \operatorname{sech}[2(z/L)]$, the PPR is

$$N = (qE_0)^{5/2} L\, (1-\varepsilon^2)^{5/4} \exp[-Z\{1-(1-\varepsilon^2)^{1/2}\}] /(4\pi^3 m) \sim (qE_0)^{5/2} L/(4\pi^3 m) \text{ as } \varepsilon \to 0$$

with $\varepsilon = m/(qE_0 L)$ and $Z = 2\pi q E_0 L^2$.

Kim and Page showed that the minimum value of $E_0$ for meaningful MAM production can be lowered significantly below the Schwinger limit by the addition of a constant magnetic field <u>B</u> parallel to the electric field <u>E.</u> In the presence of a parallel magnetic field, the PPR of charged particles per unit time and unit cross-section is modified (as derived in [5]) to,

$$N_B = (B/E_0)(qE_0)^{5/2} L\, (1-\varepsilon^2)^{3/4} \exp[-Z\{1-(1-\varepsilon^2)^{1/2}\}] \coth[\pi B/E_0 (1-\varepsilon^2)^{1/2}]/(4\pi^2 m)$$

$$\sim (B/E_0)(qE_0)^{5/2} L \coth[\pi B/E_0]/(4\pi^2 m) \text{ as } \varepsilon \to 0).$$

In the $\varepsilon \to 0$ limit we see that, $N_B = (\pi B/E_0) \coth[\pi B/E_0]\, N$.

In SI [B] = [E/c]. This means that an electric field at the Schwinger limit corresponds to a magnetic field of B = $(10^{18}$ V/M$)$ $(3 \times 10^8$ m/s$)$ = $3 \times 10^9$ T. The magnitude of this required magnetic field is on the same order as that of a magnetar! (Hence not producible presently by humans, nor likely in the long-term future!) Alternately, using present technology PPR can be enhanced by orders 10 to 100 (or greater), if the electric field (in particular, that of a laser) is pulsed with internal modulation [3].

If MAM were produced in situ, it would either be in the form of electron/positron pairs or (for sufficiently stronger electric field strength) quark/antiquark pairs. A quark/anti-quark pair will form an uncharged pion state or multiple charged/uncharged pions, if the quark pair has sufficient kinetic energy to separate sufficiently for the strong force potential interaction energy to be greater than the mass of another quark pair. Then another quark/anti-quark pair will pop into existence and a net effect can be a pair of pions of opposite charge. More likely, an electron/positron pair will pop into existence. The charged pion pairs or electron/positron pairs can be directed by external magnetic fields to produce thrust for a spacecraft.

SUMMARY

MAM production from electric fields near or above the Schwinger limit, $E_S = 1.3 \times 10^{16}$ V/cm, is nearing feasibility. MAM PPR enhancement via the addition of magnetic fields parallel to an electric field appears viable only for a B-field of at least $10^9$ T. However, MAM PPR enhancement has proved possible using pulsed electric fields near the Schwinger limit with internal modulation.

ACKNOWLEDGEMENTS

G.C. wishes to thank the TVIW organizing committee members, especially Les Johnson, for the opportunity to speak at TVIW and to attend its wide range of excellent presentations.